\documentclass[twocolumn,showpacs]{revtex4}

\usepackage{amsmath,latexsym,graphicx,subfigure,amssymb,amsfonts,mathbbol,float,times,
mathptm}

\renewcommand{\v}[1]{\mathbf{#1}}
\newcommand{\C}[1]{\mathcal{#1}}

\newcommand{\co}[1]{\C{C}_{#1}}
\newcommand{\msf}[1]{\mathsf{#1}}

\newcommand{\bl}[0]{\bullet}
\newcommand{\ad}[0]{^{\dag}}

\newcommand{\Bra}[0]{\left\langle}
\newcommand{\Ket}[0]{\right\rangle}
\newcommand{\lst}[0]{\left|}
\newcommand{\rst}[0]{\right|}
\newcommand{\wt}[1]{\widetilde{#1}}
\newcommand{\wh}[1]{\widehat{#1}}

\newcommand{\ZZ}[0]{\mathbb{Z}}

\renewcommand{\Re}[0]{\mathfrak{Re}\!~}
\renewcommand{\Im}[0]{\mathfrak{Im}\!~}

\renewcommand{\min}[0]{{\text{min}}}

\newlength{\minus}
\newcommand{\ms}[0]{\hspace{\minus}}

\begin{document}

\title{Collective excitations in circular atomic configurations, and
single-photon traps}
\author{Hanno Hammer}
\email{H.Hammer@umist.ac.uk}
\affiliation{Department of Mathematics \\ University of Manchester
Institute of Science and Technology (UMIST) \\ P.O. Box 88 \\
Manchester M60 1QD \\ United Kingdom}
%
%\date{\today}

\begin{abstract}
Correlated excitations in a plane circular configuration of identical
atoms with parallel dipole moments are investigated. The collective
energy eigenstates, which are formally identical to Frenkel excitons,
can be computed together with their level shifts and decay rates by
decomposing the atomic state space into carrier spaces for the
irreducible representations of the symmetry group $\ZZ_N$ of the
circle. It is shown that the index $p$ of these representations can be
used as a quantum number analogously to the orbital angular momentum
quantum number $l$ in hydrogen-like systems. Just as the
hydrogen~$s$-states are the only electronic wave functions which can
occupy the central region of the Coulomb potential, the quasi-particle
corresponding to a collective excitation of the atoms in the circle
can occupy the central atom only for vanishing $\ZZ_N$ quantum number
$p$. If a central atom is present, the $p=0$ state splits into two and
shows level-crossing at certain radii; in the regions between these
radii, damped quantum beats between two "extreme" $p=0$ configurations
occur. The physical mechanisms behind super- and subradiance at a
given radius are discussed. It is shown that, beyond a certain
critical number of atoms in the circle, the lifetime of the maximally
subradiant state increases exponentially with the number of atoms in
the configuration, making the system a natural candidate for a {\it
single-photon trap}.
\end{abstract}

\maketitle

\section{Introduction}

When a collection of identical atoms is located such that their mutual
distances are comparable to the wavelength of an atomic transition,
the mode structure of the electromagnetic field is altered, at least
for photon states whose energy lies in the neighbourhood of the
associated atomic transition; at the same time, level shifts and
spontaneous emission rates of the states in the atomic ensemble
undergo changes. Furthermore, the atoms in the sample become entangled
with each other as well as with the radiation field via photon
exchange. The sample then occupies collective states whose associated
spontaneous emission rates can be greater or smaller than the
single-atom decay rate and are called {\it super}- or {\it
sub}radiant, accordingly. For a collection of two-level atoms, this
phenomenon was discovered as early as 1954 by Dicke \cite{Dicke1954a}.

Super- and subradiance in the near-field regime has been studied by a
number of authors. Reviews of the subject were given in
\cite{GrossHaroche1982a,BenedictEA1996a}. The theory of subradiance in
particular was treated in a four-paper series in
\cite{CrubellierEA1985a,CrubellierEA1986a,CrubellierEA1987a,
CrubellierEA1987b}. Wigner functions, squeezing properties and
decoherence of collective states in the near-field regime have been
presented in \cite{BenedictEA1996a,BenedictCzirjak1999a,
FoeldiEA2002a}, while triggering of sub- and superradiant states was
shown to be possible in \cite{KeitelEA1992a}. Experimentally, super-
and subradiance have been observed in
\cite{PavoliniEA1985,DeVoeBrewer1996}.

In all these investigations the so-called "small-sample approximation"
played a crucial role: Here the atoms are assumed to be so close
together that they are all subject to the same phase of the radiation
field. "Closeness" here is measured on a length scale whose unit is
the wavelength $\lambda$ of the dominant atomic transition under
consideration. The small-sample approximation is then manifested by
using an interaction Hamiltonian which is independent of the spatial
location of the constituents, similar to a long-wavelength
approximation. Thus, in this approximation, super- and subradiance
appears to be a near-field phenomenon.

In this paper we perform a detailed investigation of the
simply-excited correlated states observed in a planar circular
configuration of atoms {\it beyond} the small-sample approximation:
That is to say, we determine the level shifts and decay properties of
correlated states for atomic ensembles in configurations with
arbitrary radius of the circle, not restricted to the small-sample
limit. The simply-excited correlated states so obtained turn out to be
formally identical to Frenkel excitons, e.g., in a molecular
crystal. Conceptually, however, Frenkel excitons differ from our
collective excitations in that the former are usually thought of as
being mediated by the interatomic {\it Coulomb} interaction only
\cite{Frenkel1931a,Davydov,Knox,KenkreReineker}, whereas, in the
latter case, the delocalization of the excitation energy occurs
through the exchange of {\it transverse} photons and therefore
proceeds through {\it radiative processes}. Below we shall comment on
this conceptual difference in greater detail.

Taking advantage of the discrete rotational symmetry $\ZZ_N$ of the
circle, the complex eigenvalues and eigenstates of the non-Hermitean
channel Hamiltonian governing the dynamics in the radiationless
subspace of the system can be determined analytically by
group-theoretical methods. The physical mechanism determining super-
or subradiance of a given eigenstate is then discussed. In particular
we show that each of the collective eigenstates has certain ranges of
the radius of the circle for which it can be the maximally super- or
subradiant state. We demonstrate that, for an ensemble with $N$ outer
atoms in the circle, there are $(N-1)$ states which are insensitive to
the presence of a central atom; this means that the associated
wavefunctions, level shifts and decay rates of the quasi-particle
describing the collective excitation do not change when the atom at
the center is removed, simply because the latter is not
occupied. There are only two states in a circular ensemble which
occupy the central atom; we show that they are analogues of the
$s$-state wave functions in hydrogen-like systems: Just as the $s$
states carry angular momentum quantum number $l=0$ and therefore
transform under the identity representation of $SO(3)$, our collective
$p=0$ states correspond to the identity representation of the symmetry
group $\ZZ_N$; and just as the $s$ states are the only ones which are
nonvanishing at the center-of-symmetry of the Coulomb potential in a
hydrogen-like system, so are our $p=0$ states the only quasi-particle
states which occupy the central atom at the center-of-symmetry of the
circle. These two $p=0$ states for the configuration with central atom
exhibit quantum beats between two extreme configurations, one, in
which only outer atoms are occupied, the other, in which only the
central atom is excited. At certain radii the two levels cross, and no
quantum beats occur; in this case the population transfer between the
two extreme configurations is completely aperiodic. For other radii,
the beat frequency is so much smaller than the decay rate of the
collective state as a whole, that the decay effectively proceeds like
an aperiodic "population swap". Finally we show numerically that, by
increasing the number of atoms in the circle for fixed radius, we
arrive at a domain where the minimal spontaneous decay rate in the
sample decreases exponentially with the number of atoms in the
circle. Such configurations therefore are natural candidates for {\it
single-photon traps}. This pronounced photon trapping may even survive
in the presence of dephasing interactions with an environment, or more
generally, when the system is subject to a (moderate) loss of quantum
coherence due to external noise.

\section{Spontaneous Decay and Level Shifts in a Sample of Correlated
Atoms} \label{DecayAndLevelShifts}

In this work we do not study the most general collective states in an
atomic ensemble but focus on those states which are coupled to {\bf
one} resonant photon only; we shall call these states {\it
simply-excited}. The simply-excited {\it uncorrelated} states of the
sample are given as $|e_1,g_2,\cdots,g_N,0\rangle \equiv |1,0\rangle$,
$\ldots$, $|g_1,g_2,\cdots,e_N,0\rangle \equiv |N,0\rangle$, where
$|g_A\rangle$, $|e_A\rangle$ denote the ground and excited level of
the $A$th atom, and $|0\rangle$ is the radiative vacuum. These states
are coupled to the continuum of one-photon states
$|g_1,g_2,\cdots,g_N,{\bf k}s\rangle \equiv |G, \v{k}s \rangle$, where
$({\bf k}s)$ denote the wave vector and polarization of the
photon. The uncorrelated states $|A,0\rangle$ are occupied by a sample
of atoms for which the following is true: 1.) the atoms have infinite
mutual distance; and 2.) precisely one of the $N$ atoms is excited.

If we now think of adiabatically decreasing the spatial separation of
the atoms to finite distances, the atoms will begin to "feel" each
other via their mutual Coulomb (dipole-dipole) interaction and their
coupling to the radiation field, and the states $|A,0\rangle$ will no
longer be the stationary energy eigenstates. Rather, we expect that
superpositions of uncorrelated excitations will emerge,
\begin{equation}
\label{coll1}
 \lst \C{C}\Ket = \sum_{B=1}^N c_B\, |B,0\rangle \quad,
\end{equation}
with coefficients $c_B$ which are obtained by diagonalization of a
suitable effective Hamiltonian, given below. 

Inherent in our model is the assumption that the atoms taking part in
the collective excitation have mutual distances on the order of
magnitude of an optical wavelength, i.e., several hundred
nanometers. As a consequence, electronic wavefunctions pertaining to
different atoms do not overlap, and we can refrain from
antisymmetrization of the total electronic state (Slater
determinant). This justifies the use of product states for the
simply-excited uncorrelated states $\lst A, 0\Ket$. Furthermore, it is
clear that under these conditions, the migration of excitation energy
between the different atoms can proceed only through the exchange of
transverse photons, as the Coulomb dipole-dipole interaction will play
a role only for interatomic distances $R_{AB} \ll \lambda_{eg}$, where
$\lambda_{eg}$ is the dominant wavelength of the Bohr transition. This
feature seems to set the correlated states (\ref{coll1}) apart from
the usual notion of Frenkel excitons
\cite{Frenkel1931a,Davydov,Knox,KenkreReineker}, although both are
formally identical in being superpositions of simply-excited product
states. In the case of a Frenkel exciton, however, it is usually
assumed that the atoms involved in the collective excitation are
located on the lattice sites of a crystal and therefore have mutual
distances comparable to a Bohr radius. In this case, the ``excitation
transfer interaction'' \cite{Knox} responsible for the migration of
excitation energy is Coulombic, and indeed this is assumed in all
standard accounts on Frenkel excitons (Coulomb means direct plus
exchange interaction, if relevant). To quote from a standard reference
(\cite{Davydov}, p. 19), ``Excitons ... are idealized elementary
excitations in whose consideration we ignore the delay effects and
take into account only Coulomb interaction''. This points out an
important conceptual difference between our collective excitations and
the standard notion of a Frenkel exciton.

We can express this difference in even greater detail: In the case of
the ``traditional'' Frenkel exciton, the system under consideration is
comprised of the totality of atoms in the crystal, interacting through
the instantaneous Coulomb potential or its multipolar
approximation. On account of this interaction, the atoms become
entangled with each other, but certainly remain unentangled with the
transverse degrees of freedom of the radiation field, at least as long
as only quasi-resonant atom-photon interactions (rotating-wave
approximation) are taken into account. On the other hand, the
collective states considered in our work here are states of the total
system atoms+radiation, and as a consequence, in any of these states
the atoms must be regarded as being entangled not only with each other
but also with the transverse radiative degrees of freedom. This
entanglement, of course, is a direct consequence of the fact that the
migration of excitation energy proceeds through exchange of a
transverse photon: the excited atom $A$ emits a quasi-resonant photon
which, in turn, gets absorbed by atom $B$, thereby exciting the
latter. The electrostatic interaction plays a role in this process
only for interatomic distances which are small compared to an optical
wavelength. Due to these conceptual differences, we hesitate to call
our collective states Frenkel excitons.
%----------------------------------------------------------------------
\begin{figure}[htb]
\begin{minipage}[b]{0.35\textwidth}
   \centering \includegraphics[width=1.\textwidth]{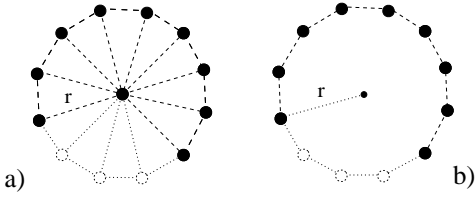}
\end{minipage}
\caption{The two circular configurations of $N(+1)$ atoms which are
examined: In configuration~a), $N$ outer atoms form a regular
$N$-polygon with radius $r$, plus an atom at the center. Configuration
b) is the same as a), but without the central
element. \label{figure1}}
\end{figure}
%----------------------------------------------------------------------

We now present the details of the computation: Let $N$ neutral atoms
be labelled by $A=1,2,\ldots, N$, each atom being localized around a
fixed point $\v{R}_A$ in space, where $\v{R}_A$ are $c$-numbers, not
operators. The atoms are assumed to be identical, infinitely heavy,
and having a spatial extent on the order of magnitude of a Bohr radius
$a_0$. The point charges within each ensemble are labelled as
$q_{A\alpha}$. On taking the long-wavelength approximation in the
minimal-coupling Hamiltonian and performing the {\it G\"oppert-Mayer
transformation} we arrive at the electric-dipole Hamiltonian
\begin{subequations}
\label{su2}
\begin{gather}
 H = \sum_A \sum_{\alpha} \frac{\v{p}_{A\alpha}^2}{ 2 \, m_{A\alpha}}
 + \sum_A \sum_{\alpha < \beta} V_{A\alpha\beta} + \sum_j
 \hbar\omega_j a_j\ad a_j - \label{su2a} \\
 - \sum_A \v{d}_A \bl \v{E}_{\bot}(\v{R}_A) = H_0 + H_I \quad,
   \label{su2b}
\end{gather}
\end{subequations}
where the unperturbed Hamiltonian $H_0$, given in (\ref{su2a}),
contains the sum over atomic Hamiltonians $H_{0A}$ including the
Coulomb interaction $V_{A\alpha\beta}$ between the internal
constituents of atoms $A= 1, \ldots, N$, but without the Coulomb
dipole-dipole interaction $V^{\text{dip}}_{AB}$ between atoms $A$ and
$B$, for $A \neq B$; and the normally-ordered radiation operators. The
electric-dipole interaction is given in (\ref{su2b}), where
$\v{E}_{\bot}$ denotes the transverse electric field operator. The
interatomic Coulomb dipole-dipole interaction $V^{\text{dip}}_{AB}$
seems to be conspicuously absent in (\ref{su2}); however, it is a
feature of the G\"oppert-Mayer transformation to transform this
interaction into a part of the transverse electric field, so that the
Coulomb interaction will emerge as a factor in the level-shift
operator, as will be seen below in
formulae~(\ref{dipodipo1}--\ref{dipodipo2}). As a consequence of being
a part of the transverse electric-field operator, the interatomic
Coulomb interaction is now fully retarded, in contrast to the
instantaneous Coulomb interaction in the minimal-coupling Hamiltonian.

The process we study in this work consists of the radiative decay of a
simply-excited correlated atomic state, or a ``Frenkel exciton'',
which is resonantly coupled to a continuum of one-photon states. For
this reason, two-photon- or higher-photon-number processes can be
expected to play a negligible role, so that it is admissible to
truncate the possible quantum states of the radiation field to
one-photon states $\lst \v{k}s\Ket$, and the vacuum $|0\rangle$. The
{\it rotating-wave approximation} is now taken \cite{Loudon}, Thus,
amongst the admissible single-photon transitions $\lst \C{C}\Ket
\rightarrow \lst G, \v{k}s\Ket$ and $\lst \C{C}\Ket \rightarrow \lst
AB, \v{k}s\Ket$, only processes of the first kind are taken into
account. The radiationless subspace $Q$ is spanned by the states $\lst
A,0\Ket$, using the same symbol for the associated projector $Q =
\sum_{A=1}^N \lst A,0\Ket \Bra A,0\rst$, while $P$ denotes the
projector onto the subspace of one-photon states $\lst G, \v{k}
s\Ket$.

\begin{figure}[htb]
\begin{minipage}[b]{0.43\textwidth}
   \centering
   \includegraphics[width=1.\textwidth]{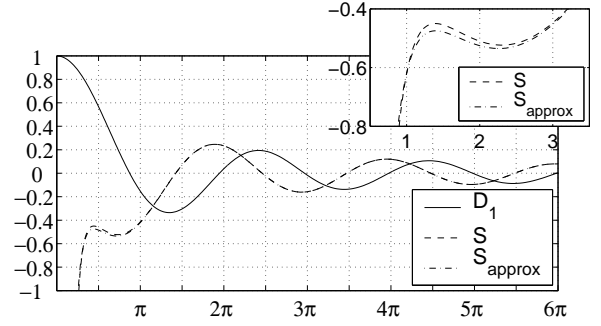}
\end{minipage}
\caption{Plots of the correlation functions $D_1(X)$ and $S(X)$
responsible for spontaneous decay (''D'') and level shifts (''S''),
respectively. $X$ is equal to $2\pi$ times the interatomic distance
measured in units of the wavelength $\lambda_{eg}$ of the dominant
atomic transition. $D_1$ becomes maximal at zero interatomic distance
while $S$ tends to minus infinity there, due to Coulomb- and radiative
self-energy of the atomic dipole. Both functions tend to zero as $X
\rightarrow\infty$, expressing the fact that any kind of correlation
must cease to exist for infinite distance. The inset shows a
comparison between the exact shift function $S(X)$, based on
eq.~(\ref{eig1b}), and the approximation $S_{\text{approx}}$, based on
eq.~(\ref{appr2}). \label{figure2}}
\end{figure}

In the present scenario we assume that the system atoms+radiation is
closed and therefore evolves unitarily; a qualitative consideration of
the impact of dephasing interactions will be given in the last
section~\ref{Dephasing}. If, at time $t=0$, the system is in a
correlated state~(\ref{coll1}), the probability of finding the
radiation at $t>0$ to be still trapped in the system is determined by
the $Q$-space amplitudes
\begin{equation}
\label{amp3}
\begin{gathered}
 \Bra A,0\rst U(t,0) \lst \C{C}\Ket = - \sum_{B=1}^N \frac{1}{2\pi i}
 \int dE\; e^{-\frac{i}{\hbar} Et}\; \cdot \\
 \cdot \Bra A,0\rst\, QG(E_+)Q\, \lst B,0\Ket\; c_B \quad,
\end{gathered}
\end{equation}
where the $Q$-space Green operator $Q G(E_+) Q$ is given in terms of
the non-Hermitean $Q$-channel Hamiltonian $\C{H}(E_+)$ as $Q G(E_+) Q
= 1/(E_+ - \C{H}(E_+))$. At the initial energy $E_i = E_e + (N-1) E_g$,
where $E_e, E_g$ are the energies of the excited and ground state of a
single atom, we have
\begin{gather}
 \C{H}(E_{i+}) = Q H_0 Q + \hbar \Delta(E_i) - i\, \frac{\hbar}{2}\,
 \Gamma(E_i) \quad,
 \label{proc2}
\end{gather}
where $\Delta(E_i)$ is the level-shift operator
\begin{subequations}
\label{eig1}
\begin{align}
 \Delta_{AB}(E_i) & = -\frac{1}{2} \Gamma \, S(X) \quad, \quad X =
 k_{eg} R_{AB} \label{eig1a} \\
 S(X) & = \frac{3}{2} \left\{ \frac{\cos X}{X} - \frac{ \sin X}{X^2} -
 \frac{ \cos X}{X^3} + \right. \nonumber \\
 & + \left. \frac{1}{\pi X^2} \int\limits_0^{\infty} du\; \frac{
 e^{-u} \left( 1 + u + u^2 \right)}{ u^2 + X^2} \right\}
 \quad. \label{eig1b}
\end{align}
\end{subequations}
and $\Gamma(E_i)$ is the decay operator
\begin{subequations}
\label{proc13}
\begin{align}
 \Gamma_{AB}(E_i) & = \Gamma\, D_1( k_{eg} R_{AB}) \quad,
 \label{proc13a} \\
 D_1(X) & = \frac{3}{2} \left( \frac{\sin X}{X} + \frac{\cos X}{X^2} -
 \frac{\sin X}{X^3} \right) \quad, \label{proc13c}
\end{align}
\end{subequations}
in the $Q$-space, where $\Gamma = d^2 k_{eg}^3/(3\pi \epsilon_0 \hbar)$
is the spontaneous emission rate of a {\it single} atom in a radiative
vacuum and $E_{eg} = E_e- E_g = \hbar c k_{eg}$. The level-shift
function $S$ can be approximated by
\begin{equation}
\label{appr2}
\begin{gathered}
  S(X)_{\text{approx}} \simeq \frac{3}{2} \left\{ \frac{\cos X}{X} -
 \frac{ \sin X}{X^2} - \frac{ \cos X}{X^3} + \right. \\
 + \left. \frac{1}{\pi X^2} \left[ \frac{1- X^2}{X}\, \arctan
 \frac{1}{X} + 1 + \frac{1}{2}\, \ln \frac{1 +X^2}{X^2} \right]
 \right\} \quad.
\end{gathered}
\end{equation}
A plot of the functions $D_1$, $S$ and $S_{\text{approx}}$ is given in
Fig.~\ref{figure2}.

[Remark: In the more general case of arbitrarily aligned atoms there
are two correlation functions $D_1, D_2$ emerging in the decay matrix
rather than just one. This is indicated by our denotation. For
parallely aligned atoms, the contribution of $D_2$ vanishes.]

A function similar to $D_1$, but within a classical context, is the
dyadic Green function of the electromagnetic field~\cite{deVriesEA1998a}.

The off-diagonal elements of the level-shift matrix $\Delta_{AB}$ can
be put into the form
\begin{equation}
\label{dipodipo1}
 \Delta_{AB}(E_i)  = \frac{d^2}{4\pi\epsilon_0 \hbar}
 \frac{1}{R_{AB}^3} \; f(R_{AB}) \quad,
\end{equation}
where $f(R_{AB})$ is an analytic function of $R_{AB}$ which tends to
$\frac{1}{2}$ as $R_{AB}$ tends to zero. Comparison with the Coulomb
dipole-dipole interaction between the two atoms $A$ and $B$,
\begin{equation}
\label{dipodipo2}
 V^{\text{dip}}_{AB} = \frac{ \v{d}_A \bl \v{d}_B - 3\, \left( \v{d}_A
 \bl \v{e}_{AB} \right) \left( \v{d}_B \bl \v{e}_{AB} \right) }{ 4\pi
 \epsilon_0 \, R_{AB}^3 }  = \frac{d^2 }{ 4\pi \epsilon_0\, R_{AB}^3}
\end{equation}
shows that the level shifts indeed contain the dipole-dipole Coulomb
interaction between the charged ensembles $A$ and $B$. On the other
hand, the diagonal elements of $\Delta_{AB}$ diverge due to Coulomb
and radiative self-energy; to remove this deficiency we renormalize
\cite{BjorkenDrell1, BjorkenDrell2, BogoliubovShirkov, ItzyksonZuber,
RyderQFT, BailinLoveGFT} the initial energy (``bare'' energy) by
redefining
\begin{equation}
\label{eig1B1a}
 E_i \longrightarrow E_i + \lim\limits_{X \rightarrow 0} \hbar
 \Delta_{AA}(E_i) \equiv \widetilde{E}_i \quad,
\end{equation}
and subtracting the same infinity from the level-shift matrix,
\begin{equation}
\label{eig1B2}
 \Delta_{AB}(E_i) \longrightarrow \widetilde{\Delta}_{AB}(E_i)
 \equiv \begin{cases} \quad \Delta_{AB}(E_i) & , \quad A \neq B
 \; , \\ \quad 0 & , \quad A=B \; , \end{cases}
\end{equation}
where $\widetilde{E}_i$ is now assumed to have a finite value, namely
the value unperturbed by the presence of $N-1$ other atoms; as a
consequence, $E_i$ must be assumed to have been infinite in the first
place. The only observable level-shifts are now those due to {\it
inter}atomic interactions.

The $Q$-channel Hamiltonian in the uncorrelated basis now becomes
\begin{subequations}
\label{eig2B}
\begin{gather}
 \Bra A,0\rst \C{H}(E_i) \lst B,0\Ket = \widetilde{E}_i\, \Eins_N -
 \frac{\hbar \Gamma}{2}\, \C{R} \quad, \label{eig2Ba} \\[7pt]
 \C{R} = \left(
 \begin{array}{*3{c}} i & S+iD_1 & \cdots \\ S+iD_1
 & i & \cdots \\ \vdots &\vdots & \ddots \end{array} \right) = \C{R}^T
 \quad. \label{eig2Bb}
\end{gather}
\end{subequations}
This matrix can be diagonalized in terms of left and right
eigenvectors \cite{MorseFeshbach1+2}
\begin{equation}
\label{eig3A}
\begin{gathered}
 \Bra A,0 \rst \C{H}(E_i) \lst B,0\Ket = \\\
 = \sum_{p=1}^N \Bra A,0\rst \left. \co{p} \Ket \bigg\{ \wt{E}_i
 -\frac{\hbar \Gamma(k_{eg})}{2}\,\mu_p \bigg\} \Bra \co{p}^* \rst
 \left. B,0\Ket \quad.
\end{gathered}
\end{equation}
where
\begin{subequations}
\label{eig6}
\begin{align}
 \wt{\Delta}_p & = -\frac{\Gamma}{2}\; \Re\mu_p \quad,
 \label{eig6a} \\
 \Gamma_p & = \ms \Gamma\; \Im\mu_p \quad, \label{eig6b}
\end{align}
\end{subequations}
%----------------------------------------------------------------------
%----------------------------------------------------------------------
For the given radius $r$, number of atoms $N$, and configuration (a)
or (b), let $\lst \C{C}_{\min} \Ket$ be the eigenstate with the
smallest decay rate,
\begin{equation}
\label{amp10}
 \Gamma_\min \le \Gamma_p \quad \text{for all $p=1, \ldots, N$} \quad.
\end{equation}
Then the associated correlated state
\begin{equation}
\label{amp11}
 \lst \C{C} \Ket = \frac{1}{\sqrt{\Bra \co{\min} | \co{\min} \Ket}}\,
 \lst \co{\min} \Ket \quad
\end{equation}
is the most stable one with respect to spontaneous decay [saying
nothing about stability against environmental perturbations], and
hence is a candidate for a single-photon trap.

\section{Cyclic symmetry of the circular atomic configurations}

We now construct the eigenvectors $\lst \co{p} \Ket$ of $\C{H}(E_i)$
explicitly by group-theoretical means, taking advantage of the fact
that the system has a cyclic symmetry group
\begin{subequations}
\label{cyclic1}
\begin{align}
 G & = \left\{ e, T, T^2, \ldots, T^{N-1} \right\} \quad,
 \label{cyclic1a} \\
 T^{N} & = e \quad, \label{cyclic1b}
\end{align}
\end{subequations}
where $N$ is the number of outer atoms along the perimeter of the
circle, the generator $T$ acts on coordinate space as a (passive)
rotation by the angle $2\pi/N$ about the symmetry axis of the circle,
and acts on the state space unitarily by relabelling the outer atoms,
but leaving the central atom invariant,
\begin{equation}
\label{cyclic12}
\begin{aligned}
 T\, \lst A\Ket & = \lst A-1 \Ket \quad, \\
 T\, \lst z\Ket & = \lst z\Ket \quad.
\end{aligned}
\end{equation}
The {\it unitary irreducible} representations of $\ZZ_N$ are given by
\begin{equation}
\label{cyclic2}
 \Gamma^p\left( T^A \right) = \exp\left(\frac{2\pi i p A}{N} \right)
 \quad, \quad A = 1,2, \ldots, N \quad,
\end{equation}
and using appropriate projection operators, the radiationless
$Q$-space can be decomposed into an orthogonal direct sum $Q =
\oplus_{p=0}^{N-1} \msf{T}_p$ of carrier spaces $\msf{T}_p$ of
irreducible representations $\Gamma^p$. Since the channel Hamiltonians
of both configurations are invariant under this transformation,
\begin{equation}
\label{cyclic18}
 T^{-1}\, \C{H}\, T = \C{H} \quad,
\end{equation}
$\C{H}$ preserves all carrier spaces $\msf{T}_p$ and hence can be
diagonalized on each $\msf{T}^p$ separately. This simplification makes
it possible to perform the diagonalization analytically.

\section{Eigenspaces of the generator $T$ of the symmetry group}

\begin{figure}[htb]
\begin{minipage}[b]{0.3\textwidth}
   \centering
   \includegraphics[width=1.\textwidth]{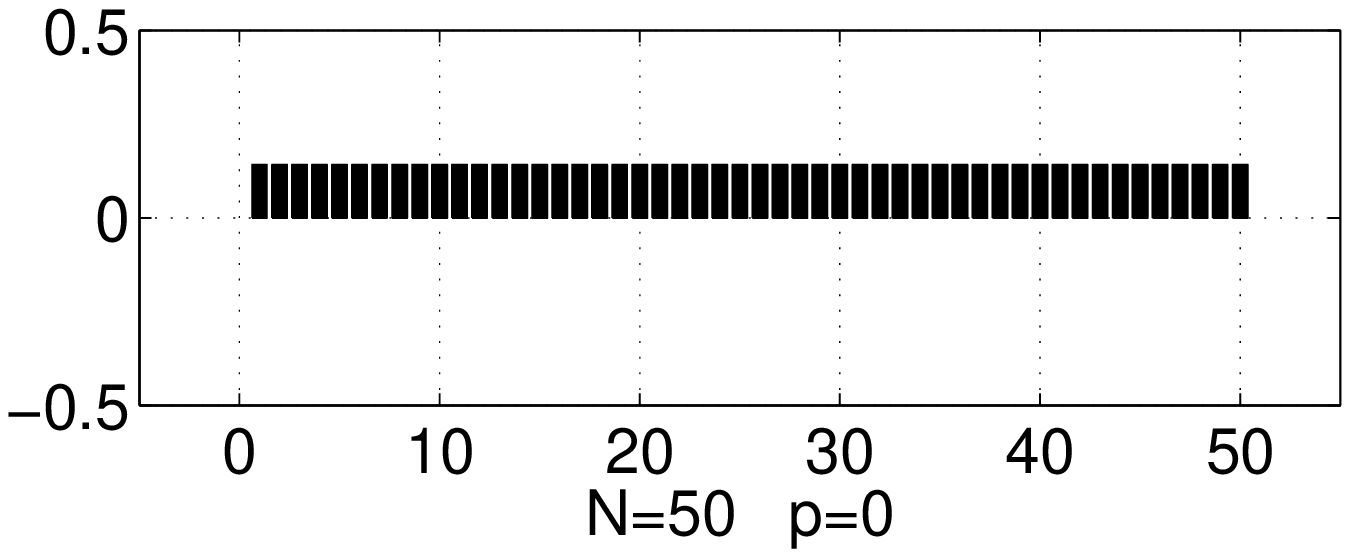}
\end{minipage} \\
\begin{minipage}[b]{0.3\textwidth}
   \centering
   \includegraphics[width=1.\textwidth]{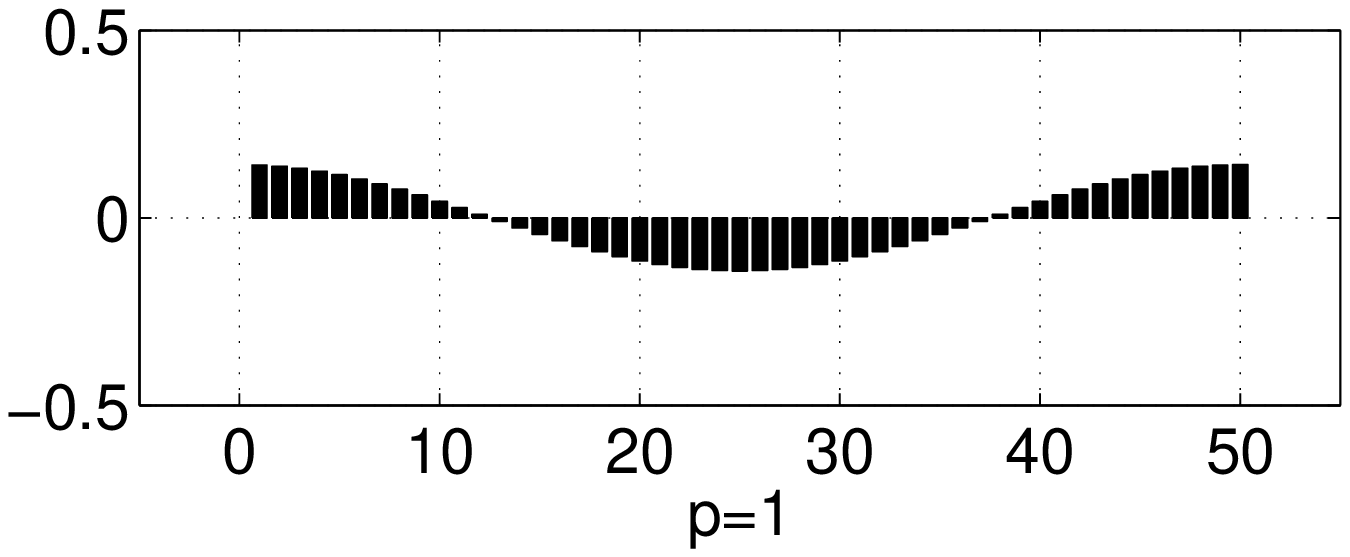}
\end{minipage} \\
\begin{minipage}[b]{0.3\textwidth}
   \centering
   \includegraphics[width=1.\textwidth]{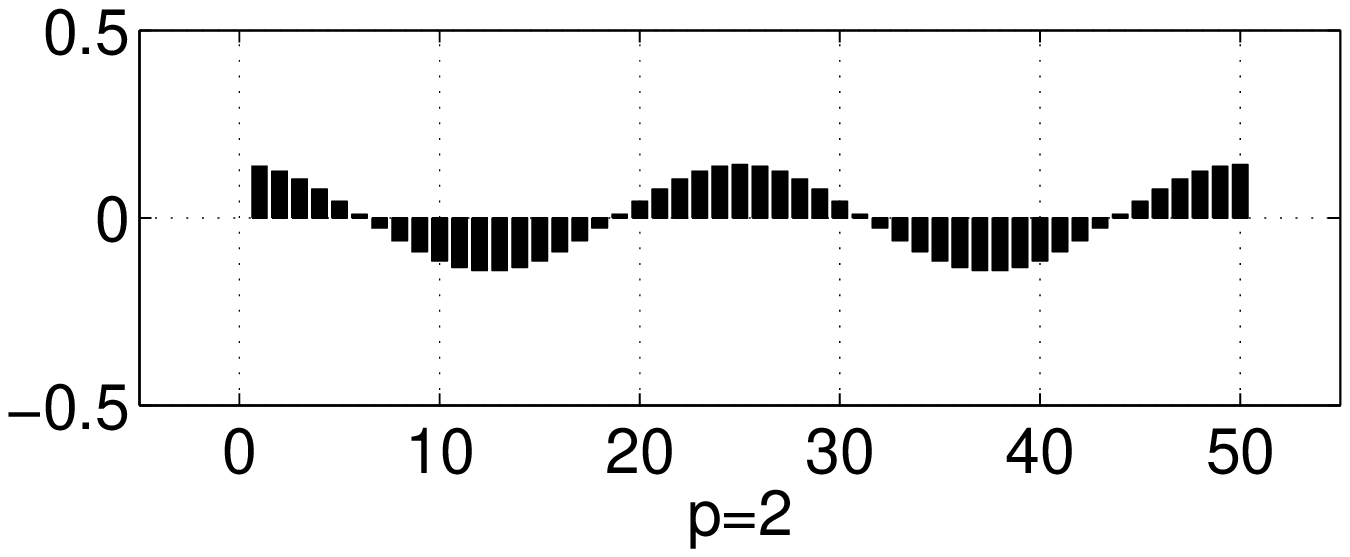}
\end{minipage} \\
\begin{minipage}[b]{0.3\textwidth}
   \centering
   \includegraphics[width=1.\textwidth]{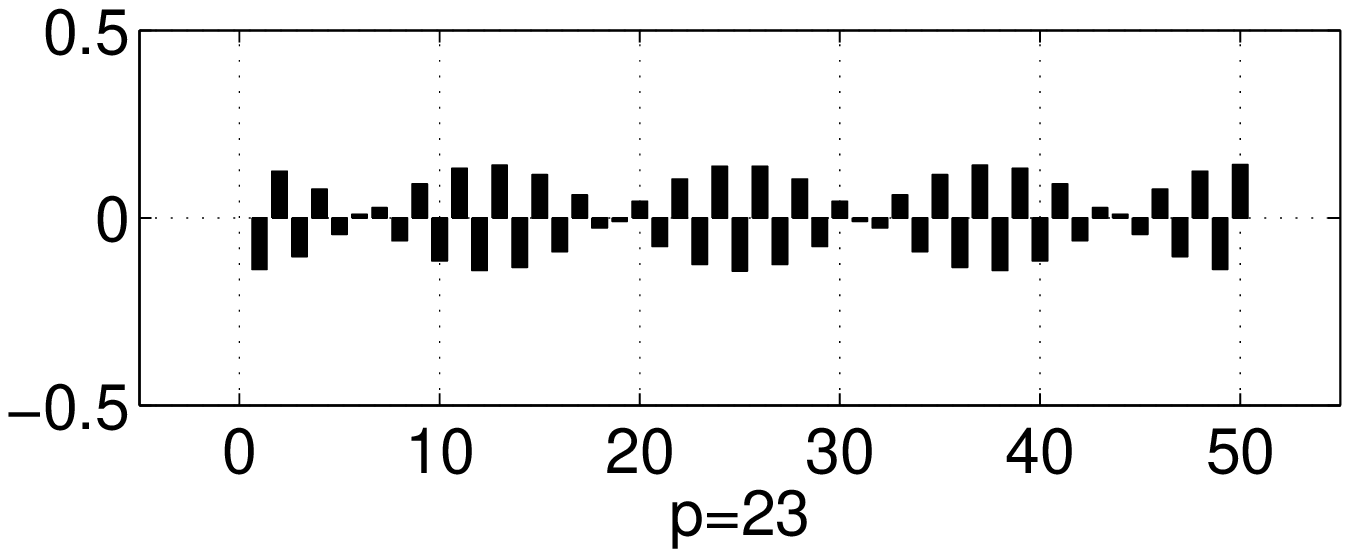}
\end{minipage} \\
\begin{minipage}[b]{0.3\textwidth}
   \centering
   \includegraphics[width=1.\textwidth]{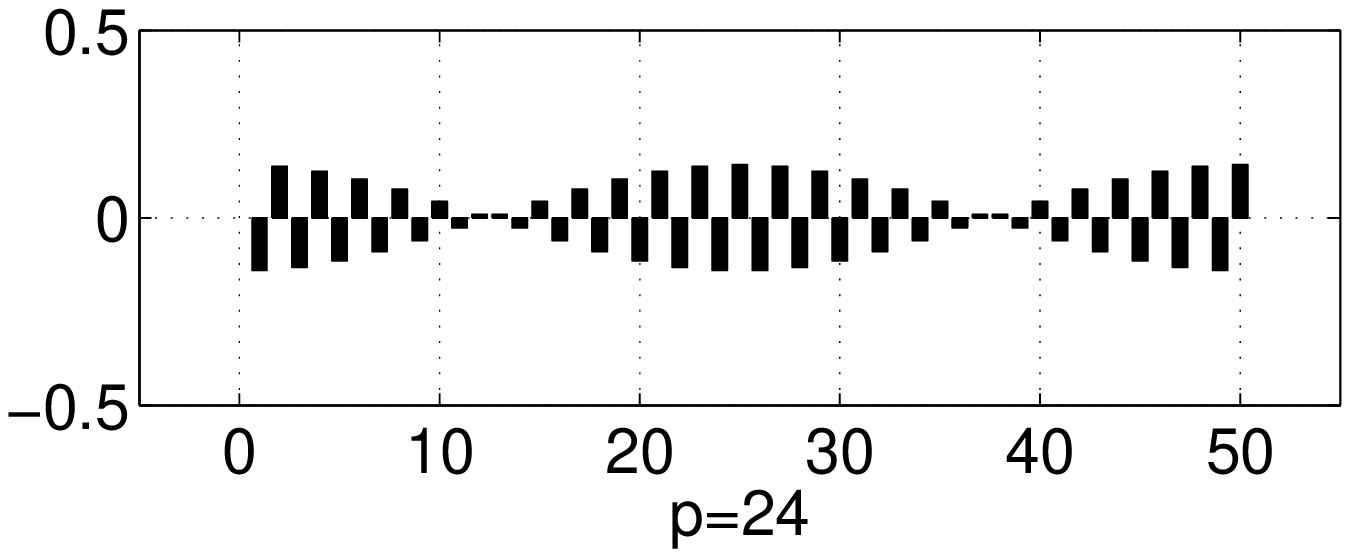}
\end{minipage}
\begin{minipage}[b]{0.3\textwidth}
   \centering
   \includegraphics[width=1.\textwidth]{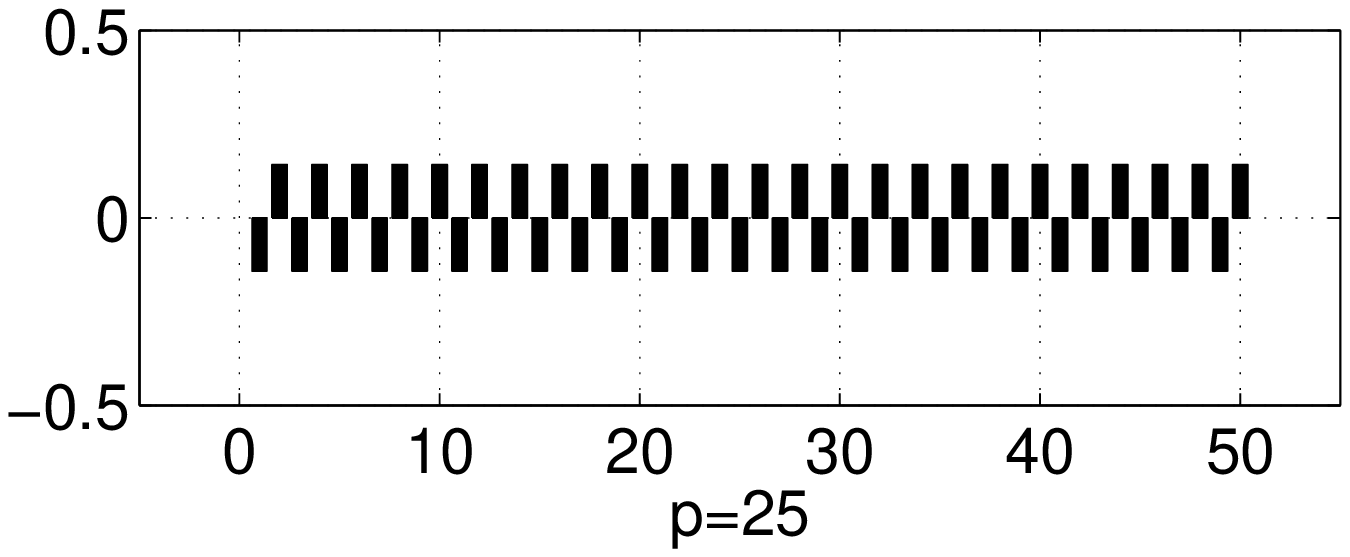}
\end{minipage}
\caption{The real parts $\Re c^p_A$ of the components of states
(\ref{eigen8a}, \ref{eigen7}) in the uncorrelated basis, for the
eigenstates $p=0,1,2,23,24,25$ of a configuration with $N=50$ outer
atoms. For $p \neq 0$ these coefficients are the same for both
configurations (a) and (b). \label{figure3}}
\end{figure}

In order to perform the diagonalization we need to construct basis
vectors of the carrier spaces $\msf{T}_p$ by applying the standard
projection operators \cite{Cornwell1} $\C{P}^p$,
\begin{equation}
\label{eigen2a}
 \C{P}^p = \frac{1}{N} \sum_{A=0}^{N-1} \exp\left( -\frac{2\pi i p
 A}{N} \right)\, T^A \quad,
\end{equation}
projecting onto $\msf{T}_p$, on an arbitrary correlated state $\lst
 \co{} \Ket = c_z |z,0\rangle + \sum_{A=1}^N c_A |A,0\rangle$, with
 the result
\begin{equation}
\label{eigen6}
 \C{P}^p \lst \co{} \Ket = \delta_{p0}\, c_z\, \lst z,0 \Ket + c
 \sum_{A=1}^N \exp \left( \frac{2\pi i p A}{N} \right)\, \lst A, 0\Ket
 \quad,
\end{equation}
where $|c_z|^2 + |c|^2 = 1$. It follows that, for $p \neq 0$, the
carrier spaces $\msf{T}_p$ are one-dimensional, and are spanned by
normalized basis vectors
\begin{equation}
\label{eigen7}
 \lst \co{p} \Ket \equiv \frac{1}{\sqrt{N}} \sum_{A=1}^N \exp \left(
 \frac{2\pi i p A}{N} \right)\, \lst A, 0\Ket \quad,
\end{equation}
which turn out to be simple Fourier transforms of uncorrelated
excitations. This statement is true for both configurations (a) and
(b). We emphasize again that, here, $N$ denotes the number of {\it
outer} atoms. In Fig.~\ref{figure3} we plot the real parts of $c^p_A =
\Bra A, 0 | \C{C}_p \Ket$ for some of the $p$-states with $N=50$ outer
atoms.
%----------------------------------------------------------------------
%----------------------------------------------------------------------

If $p=0$, the associated eigenspace $\msf{T}_0$ is one-dimensional for
configuration (b), and is spanned by
\begin{subequations}
\label{eigen8}
\begin{equation}
\label{eigen8a}
 \lst \co{0} \Ket \equiv \frac{1}{\sqrt{N}} \sum_{A=1}^N \lst A, 0\Ket
 \quad,
\end{equation}
while, for configuration (a), $\msf{T}_{0}$ is two-dimensional, with a
second basis vector
\begin{equation}
\label{eigen8b}
 \lst \co{z} \Ket \equiv \lst z, 0\Ket \quad.
\end{equation}
\end{subequations}
The two states in~(\ref{eigen8}) are totally isotropic under the
symmetry group $\ZZ_N$ in the sense that they do not pick up a phase
factor when we operate with a group element on them. Furthermore, they
are ``extreme'' states, since in~(\ref{eigen8a}), only outer atoms are
occupied, with all atomic dipoles aligned parallely, as can be seen in
the first plot in Fig.~\ref{figure3}; while in~(\ref{eigen8b}), only
the central atom is occupied.

\section{Diagonalization of the channel Hamiltonian on carrier spaces
 of the symmetry group}

\subsection{Configuration (b)}

%----------------------------------------------------------------------
\begin{figure}[htb]
%\vspace{1em}
\begin{minipage}[b]{0.45\textwidth}
   \centering \includegraphics[width=1.\textwidth]{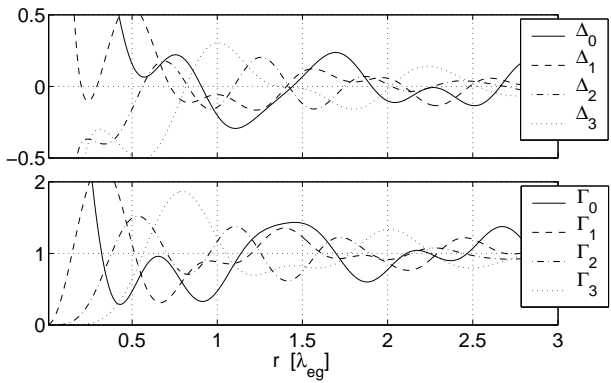}
\end{minipage}
\caption{Relative level shifts $\wt{\Delta}_p /\Gamma$ and decay rates
$\Gamma_p / \Gamma$ for the states $p=0,1,2,3$ with $N=7$ outer atoms
and no central atom. From the second plot we see that none of the
$p$-states can generally be declared as super- or subradiant; rather,
for each state there exist ranges of the radius $r$ for which the
state $\lst \co{p}\Ket$ is maximally super- or maximally
subradiant. The physical reason for this dependence lies in the
interference due to varying phase differences between spatially
retarded radiative contributions from all but one atom, at the site of
the given atom. \label{figure4}}
\end{figure}
%----------------------------------------------------------------------
We first discuss configuration (b) without central atom. In this case,
each of the carrier spaces $\msf{T}_p$ is one-dimensional, and is
spanned by states~(\ref{eigen7}, \ref{eigen8a}). Since $\C{H}$ and
$\C{R}$ preserve these carrier spaces it follows that each of the
$\lst \co{p}\Ket$, $p=0, \ldots, N-1$ is automatically a right
eigenvector of $\C{R}$, with eigenvalue
\begin{subequations}
\label{diago11}
\begin{align}
 \mu_p & = - \frac{2 \wt{\Delta}_p}{\Gamma} + i
 \frac{\Gamma_p}{\Gamma} \quad, \label{diago11a} \\
 \wt{\Delta}_p & = - \frac{\Gamma}{2} \sum_{A=2}^N S(k_{eg} R_{1A})\,
 \cos \left( \frac{2\pi p (A-1) }{N} \right) \quad, \label{diago11b}
 \\
 \Gamma_p & = \Gamma \Bigg\{ 1 + \sum_{A=2}^N D_1(k_{eg} R_{1A})\,
 \cos \left( \frac{2\pi p (A-1) }{N} \right) \Bigg\} \quad,
 \label{diago11c}
\end{align}
\end{subequations}
where we have used~(\ref{append1}). It follows that
\begin{equation}
\label{diago12}
 \mu_{N-p} = \mu_p \quad,
\end{equation}
hence some of the eigenvalues are degenerate; the exact result is
presented in Table~\ref{TabDegeneracy}. Eqs.~(\ref{diago11b})
and~(\ref{diago11c}) contain the level shifts and decay rates, which
are given as cosine transformations of the correlation functions $S$
and $D_1$. A plot of these quantities for the first four eigenvalues
for $N=7$ outer atoms is given in Fig.~\ref{figure4}, where the
approximated shift function $S_{\text{approx}}$ based on
eq.~(\ref{appr2}) has been used.

\subsection{Mechanisms for super- and subradiance}

The Figures~\ref{figure3} and~\ref{figure4} give some insight into the
mechanism of super- and subradiance. We first note that the system
atoms+radiation evolves unitarily in the present treatment, and since
the initial state was pure, the state of the total system is always
pure. This fact entitles us to visualize the atoms in the sample as
being radiating {\it coherently}. This coherence is at the heart of
super- and subradiance since it means that the radiative contributions
emitted by the individual atoms will {\it interfere}: generally
speaking, a correlated state will be super- or subradiant when the
total interference between these contributions is constructive or
destructive in the far zone (measured in units of $\lambda_{eg}$) of
the radiation field around the configuration. Based on this reasoning
we now give a qualitative explanation for the behaviour of level
shifts and decay rates in Fig.~\ref{figure4} in the small-sample limit
$r \rightarrow 0$:
%------------------------------------------------------------------------------
\begin{table}
\begin{center}
{ Level degeneracy for configurations (a) and (b)}
\end{center}
\begin{equation*}
\begin{array}{| @{\hspace{.5em}} r @{\hspace{.5em}} || @{\hspace{.5em}}
r@{}c@{}l @{\hspace{.1em}} @{\hspace{.5em}} | @{\hspace{.5em}} r@{}c@{}l
@{\hspace{.5em}} |} \hline
     & N & = & 2n+1 & N & = & 2n \\ \hline
 \hline
 \text{degenerate} & p & = & \begin{array}{r@{}c@{}l} 1, & \ldots, & n
 \\ 2n, & \ldots, & n+1 \end{array} & p & = & \begin{array}{r@{}c@{}l}
 1, & \ldots, & n-1 \\ 2n-1, & \ldots, & n+1 \end{array} \\ \hline
 \text{non-degenerate} & p & = & 0\, (\pm) & p & = & 0\, (\pm), n
 \\\hline
\end{array}
\end{equation*}
\caption{Degeneracy of the eigenvalues for quantum numbers $p=0 (\pm),
\ldots, N-1$, for odd and even $N$, for configurations (a) and
(b). Vertically stacked $p$ values in the second row are mutually
degenerate. \label{TabDegeneracy} }
\end{table}
%------------------------------------------------------------------------------

We see in Fig.~\ref{figure4} that the $p=0$ state is the only one
which has a nonvanishing decay rate in this limit: All atomic dipoles
are aligned parallely in this state [see eq.~(\ref{eigen8a}) and first
plot in Fig.~\ref{figure3}], and vanishing interatomic distance on a
length scale of a wavelength means that the radiation emitted
coherently by the sample must interfere completely
constructively. Hence, the sample radiates faster than a single atom
by a factor $N$, since the decay rate is
\begin{equation}
\label{Decay0}
 \lim\limits_{r \rightarrow 0} \Gamma_{p=0} = N \cdot \Gamma \quad,
\end{equation}
as follows from eq.~(\ref{diago11}). Conversely, the suppression of
spontaneous decay for the states with $p>0$ is a consequence of the
fact that the dipoles have alternating orientations, see
eq.~(\ref{eigen7}) and the second to fifth plot in Fig.~\ref{figure3},
so that, in the small-sample limit $r \rightarrow 0$, roughly one-half
of the atoms radiate in phase, while the other half has a phase
difference of $\pi$; hence
\begin{equation}
\label{DecayP}
 \lim\limits_{r \rightarrow 0} \Gamma_{p>0} = 0 \quad.
\end{equation}
This behaviour is exemplified by the plots of $\Gamma_1, \Gamma_2,
\Gamma_3$ in Fig.~\ref{figure4}. However, the destructive interference
in the $(p>0)$ states is independent of how adjacent dipoles are
oriented, as long as $r \ll \lambda_{eg}$; in this limit, the decay
rate should not be noticeably affected by rearranging the dipoles in
different patterns as long as the 50:50 ratio of parallel-antiparallel
dipoles is kept fixed. What {\it will} be affected by such a
redistribution is the level shift, since the Coulomb interaction
between the dipoles in the sample may change towards more attraction
or repulsion between the atoms. By this mechanism we can explain the
divergent behaviour of the level shifts in the small-sample limit: The
shift of the $p=0$ state always behaves like
\begin{equation}
\label{Shift0}
 \lim\limits_{r \rightarrow 0} \Delta_{p=0} = + \infty \quad,
\end{equation}
which arises from the Coulomb repulsion of the parallely aligned
dipoles. On the other hand, the level shifts for the $(p>0)$ states
depend on the relative number of parallely aligned dipoles in the
immediate neighbourhood of a given dipole, or conversely, on the
degree of balancing the Coulomb repulsion by optimal pairing of
antiparallel dipoles. As a consequence, states for which $p$ is close
to zero always have positive level shifts, since the Coulomb
interaction between adjacent dipoles tends to be repulsive, as seen in
the first three plots in Fig.~\ref{figure3}, and the behaviour of
$\Delta_0, \Delta_1$ in Fig.~\ref{figure4}. On the other hand, states
for which $p$ is close to $N/2$ tend to have antiparallel orientation
between adjacent dipoles, hence the Coulomb interaction is now largely
attractive, which explains why these states have negative level shifts
in the limit $r \ll \lambda$. This is seen in the last three plots in
Fig.~\ref{figure3} and the behaviour of $\Delta_2, \Delta_3$ in
Fig.~\ref{figure4}.

The same mechanism clearly also governs the super- or subradiance of
the sample with finite interatomic distance. In this case the
information about the orientation of surrounding dipoles at sites
$\v{R}_A$ is contained in the transverse electric field, which arrives
at the site $\v{R}_1$ with a retardation $| \v{R}_A - \v{R}_1 |/
c$. Thus, in addition to the phase difference imparted by the
coefficients $c^p_A$, there is another contribution to the phase from
the spatial retardation, which accounts for the dependence of the
level shifts and decay rates on the radius $r$. However, apart from
this additional complication, the physical mechanism determining
whether a given state is super- or subradiant is clearly the same as
in the small-sample limit, and can be traced back to the mutual
interference of the radiation emitted by each atom, arriving at a
given site $\v{R}_1$.

\subsection{Configuration (a)}

Now we turn to compute the complex energy eigenvalues and eigenvectors
for configuration (a) with central atom. For $p = 1, \ldots, N-1$, the
basis vectors carrying irreducible representations of the symmetry
group are the same as before, and are given in
eq.~(\ref{eigen7}). Consequently, they are also right eigenvectors of
the matrix $\C{R}$; the associated eigenvalues $\mu_p$ turn out to be
the same as for configuration (b) and thus are given by
formulas~(\ref{diago11}). This result means that the presence or
absence of the central atom is not ``felt'' by the system in the modes
$\lst \co{p} \Ket$, $p = 1, \ldots, N-1$, since the central atom is
not occupied in these states.

On the other hand, the eigenspace $\msf{T}_0$ of $T$ corresponding to
the $p=0$ representation is now two-dimensional and is spanned by
$\Gamma^0$-basis vectors (\ref{eigen8a}) and (\ref{eigen8b}). We now
need to diagonalize the channel Hamiltonian on this subspace: the
diagonalization of the matrix $\C{R}$ in the basis $\lst \co{0}\Ket,
\lst \co{z} \Ket$ yields
\begin{subequations}
\label{cona7B}
\begin{align}
 \lst \co{0+} \Ket & = \hspace{0.5ex} \cos\wh{\theta}\, \lst
 \co{0} \Ket + \sin \wh{\theta}\, \lst \co{z} \Ket \quad,
 \label{cona7Ba} \\
 \lst \co{0-} \Ket & = - \sin \wh{\theta}\, \lst \co{0} \Ket +
 \cos \wh{\theta}\, \lst \co{z} \Ket \quad,
 \label{cona7Bb}
\end{align}
\end{subequations}
where
\begin{equation}
\label{cona9comment}
 \wh{\theta} = \frac{i}{2} \ln \frac{i+c}{i- c} \quad,
\end{equation}
with
\begin{equation}
\label{cona9}
 c = \frac{ 2\sqrt{N}\, M(k_{eg} r) }{ \sum_{A=2}^N M(k_{eg} R_{1A}) +
 \sqrt{ \left[ \sum_{A=2}^N M( k_{eg} R_{1A}) \right]^2 + 4 N\,
 M^2(k_{eg} r)}} ,
\end{equation}
and $M = S+iD_1$. The eigenvectors (\ref{cona7B}) are {\it not}
orthogonal, consistent with the fact that the matrix $\C{R}$ is not
Hermitean. However, (\ref{cona7B}) form an orthonormal system together
with the left eigenvectors
\begin{equation}
\label{cona9B}
\begin{aligned}
 \Bra \co{0+}^* \rst & = \hspace{0.5ex} \cos\wh{\theta}\, \Bra
 \co{0} \rst + \sin\wh{\theta}\, \Bra \co{z} \rst \quad, \\
 \Bra \co{0-}^* \rst & = - \sin\wh{\theta}\, \Bra \co{0} \rst +
 \cos\wh{\theta}\, \Bra \co{z} \rst \quad.
\end{aligned}
\end{equation}
The associated eigenvalues $\mu_{0\pm}$ (of $\C{R}$) are
\begin{equation}
\label{cona13}
\begin{aligned}
 \mu_{0\pm} & = i + \frac{1}{2} \sum_{A=2}^N M\left( k_{eg} R_{1A}
 \right) \pm \\
 & \pm \frac{1}{2} \sqrt{ \left[ \sum_{A=2}^N M \left( k_{eg} R_{1A}
 \right) \right]^2 + 4 N\, M^2\left( k_{eg} r \right) } \quad,
\end{aligned}
\end{equation}
where $r$ is the radius of the circle. From (\ref{cona13}) we obtain
the level shifts $\wt{\Delta}_{0\pm}$ and decay rates $\Gamma_{0\pm}$
of the $p=0$ states as in eq.~(\ref{diago11}).

The definition of eigenvalues $\mu_{0\pm}$ as given in
eq.~(\ref{cona13}) is not yet the final one, however. In
(\ref{cona13}) we have assigned $\mu_{0+}$ to the square root with
positive real part. From (\ref{eig6a}) we see that the $\mu_{0+}$ so
defined always comes with a negative level shift, and therefore the
associated real energy $\wt{E}_i + \hbar \wt{\Delta}_{0+}$ is always
smaller than the energy associated with $\mu_{0-}$. This state of
affairs would be acceptable as long as the levels would never cross;
but crossing they do, as can be seen in Fig.~\ref{figure5}:
%----------------------------------------------------------------------
\begin{figure}[htb]
%\vspace{1em}
\begin{minipage}[b]{0.47\textwidth}
   \centering \includegraphics[width=1.\textwidth]{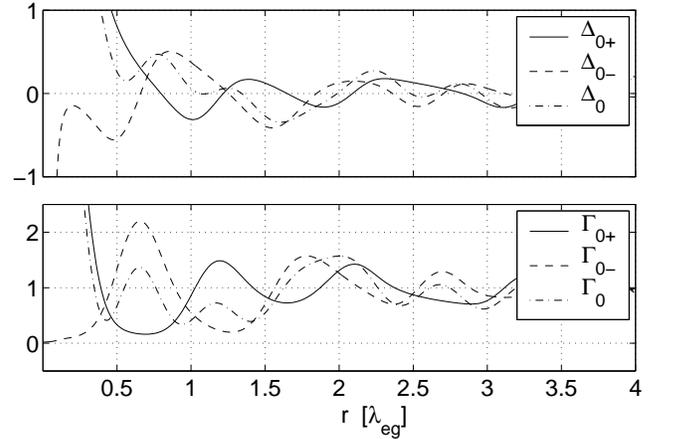}
\end{minipage}
\caption{Plotted are relative quantities $\wt{\Delta}_{0(\pm)}/
\Gamma$ and $\Gamma_{0(\pm)}/ \Gamma$. For $N=10$ outer atoms, the
level shifts $\wt{\Delta}_{0\pm}$ exhibit level crossing twice on
every wavelength, for $r \ge \lambda_{eg}$. At the radius $r$ of a
crossing, the beat frequency~(\ref{rabi6}) vanishes, preventing any
oscillatory population transfer between $\lst \C{C}_0 \Ket$ and $\lst
\C{C}_z \Ket$. The states $\lst \C{C}_{0\pm} \Ket$ are defined such
that, as $r \rightarrow 0$, the level shifts $\wt{\Delta}_{0\pm}$ tend
to $\pm \infty$, respectively. --- For comparison, level shifts and
decay rates of the state $\lst \C{C}_0 \Ket$ without central atom are
included. \label{figure5}}
\end{figure}
%----------------------------------------------------------------------
For radii greater than a certain lower bound $r_0$, which depends on
the number $N$ of outer atoms, we see two level-crossings per unit
wavelength. For $N=10$ this radius is roughly $r_0 \sim \lambda$. At
each crossing we have to reverse the assignment of square roots to
eigenvalues in order to obtain smooth eigenvalues. We therefore have
to redefine $\mu_{0\pm}$, and the associated eigenvectors, in order to
take account of this reversal. A look at Fig.~\ref{figure5} shows
that, for $r\le 0.7~\lambda$, no crossings occur, so that we can
uniquely identify the eigenvalues by their behaviour in the limit $r
\rightarrow 0$. Thus, we finally define $\mu_{0\pm}$ to be that
eigenvalue whose associated level shift $\wt{\Delta}_{0\pm} = - \Gamma
\Re \mu_{0\pm}/2$ tends towards $\pm \infty$,
respectively. Physically, the dipoles in the state $\lst \co{0+} \Ket
/ \left\| \co{0+} \right\|$ are aligned parallely, which in the limit
$r \rightarrow 0$ produces Coulomb repulsion, and hence the positive
level shift. It follows that the state $(p=0+)$ is the natural
analogue of the $(p=0)$ state in configuration (b), since they both
have the same behaviour at small radii. It is also expected to decay
much faster than a single atom, an expectation which is confirmed by
the behaviour of the imaginary part $\Im \mu_{0+} =
\Gamma_{0+}/\Gamma$, which tends to $\sim N+1$ as $r \rightarrow
0$. Again this follows the pattern of the $p=0$ state in configuration
(b). A visual comparison between states $(p=0+)$ and $(p=0)$ is given
in Fig.~\ref{figure5} for $N=10$ outer atoms.

On the other hand, in the state $\lst \co{0-} \Ket / \left\| \co{0-}
\right\|$, the central atom is now oriented antiparallelly to the
common orientation of the outer dipoles, and, in the limit $r
\rightarrow 0$, is much stronger occupied than the outer atoms, as
follows from eq.~(\ref{cona7Bb}). Hence, as a consequence of Coulomb
attraction between the outer atoms and the central atom, the energy is
shifted towards $- \infty$, and at the same time, the system has
become extremely stable against spontaneous decay: This is reflected
in the fact that, as $r \rightarrow 0$, the decay rate $\Gamma_{0-}$
tends to zero as well.

As mentioned above, the states with higher quantum numbers $p=1,
\ldots, N-1$ are the same as in configuration (b), and have the same
eigenvalues. The two levels $p=(0\pm)$ are non-degenerate, except for
accidental degeneracy, and also are non-degenerate with the $(p>0)$
levels. As a consequence, the level degeneracy is similar to case (b),
and is again expressed in Table~\ref{TabDegeneracy}.

\subsection{Quantum beat between unperturbed $p=0$ states}

As follows from eqs.~(\ref{cona7B}), the true eigenstates $\lst
\co{0\pm} \Ket$ are in general linear combinations of the
"unperturbed" irreducible basis vectors $\lst \co{0} \Ket$,
eq.~(\ref{eigen8a}), and $\lst \co{z} \Ket$, eq.~(\ref{eigen8b}); as a
consequence, the true eigenstates will give rise to a quantum beat
between $\lst \co{0} \Ket$ and $\lst \co{z} \Ket$. In order to
determine the beat frequency we compute the probability of finding the
system at time $t$ in the extreme state $\lst \C{C}_0 \Ket$ if
initially it was in the other extreme state $\lst \C{C}_z \Ket$,
\begin{align}
 P_{\lst \co{z} \Ket \rightarrow \lst \co{0} \Ket}(t) & = \Big| \Bra
 \co{0} \rst U(t,0) \lst \co{z} \Ket \Big|^2 = \left|
 \sin\wh{\theta} \cos\wh{\theta} \right|^2\, \times \nonumber \\
 & \times\, \Bigg\{ e^{- \Gamma_{0+} t} + e^{- \Gamma_{0-} t} - 2\,
 e^{-\frac{1}{2} \left( \Gamma_{0+} + \Gamma_{0-} \right) t } \times
 \nonumber \\
 & \times \cos \left[ \left( \wt{\Delta}_{0+} - \wt{\Delta}_{0-}
 \right) t \right] \Bigg\} \quad. \label{rabi4}
\end{align}
The last line shows that the oscillation between the two states occurs
at the beat frequency
\begin{equation}
\label{rabi5}
 \omega_R  = \left| \wt{\Delta}_{0+} - \wt{\Delta}_{0-} \right| \quad.
\end{equation}
The beat frequency can be expressed in terms of the function $M= S+
iD_1$ as
\begin{equation}
\label{rabi6}
 \omega_R = \frac{\Gamma}{2} \Re \sqrt{ \left[ \sum_{A=2}^N M \left(
 k_{eg} R_{1A} \right) \right]^2 + 4 N\, M^2\left( k_{eg} r \right) }
 \quad.
\end{equation}
%----------------------------------------------------------------------
\begin{figure}[htb]
%\vspace{1em}
\begin{minipage}[b]{0.4\textwidth}
   \centering \includegraphics[width=1.\textwidth]{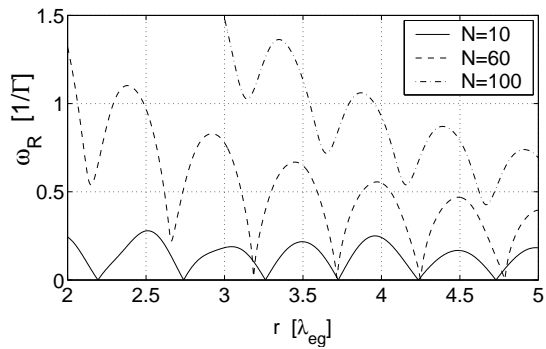}
\end{minipage}
\caption{The beat frequency eq.~(\ref{rabi6}) for three different
values $N=10,60,100$ of outer atoms, for radii between $2$ and $5$
wavelengths. At the radius of a level crossing $\wt{\Delta}_{0+} =
\wt{\Delta}_{0-}$, the beat frequency vanishes. It can be seen that
the greater the number of outer atoms $N$, the greater becomes the
radius $r_0$ beyond which level crossing occurs.
\label{figure6}}
\end{figure}
%----------------------------------------------------------------------
For a given number of outer atoms, there can exist discrete radii at
which the beat frequency $\omega_R$ vanishes. From
formula~(\ref{rabi5}) we see that this is the case precisely when the
two levels cross, and hence the normalized states $\lst \C{C}_{0+}
\Ket$ and $\lst \C{C}_{0,-} \Ket$ are degenerate in energy. This is
demonstrated in Fig.~\ref{figure6}.

\begin{figure}[htb]
\begin{minipage}[b]{0.4\textwidth}
   \centering \includegraphics[width=1.\textwidth]{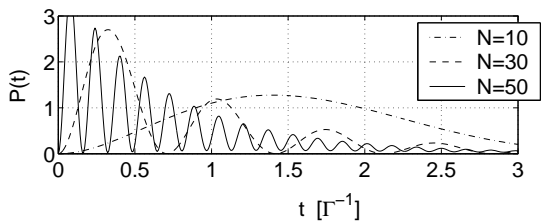}
\end{minipage}
\caption{The probability $P(t)/ | \sin \wh{\theta} \cos \wh{\theta}
  |^2$ for the transition $\lst \co{z} \Ket \rightarrow \lst \co{0}
  \Ket$, as given in eq.~(\ref{rabi4}), for $N=10,30,50$ outer atoms,
  at a radius $r= 0.45~\lambda_{eg}$. The beat oscillations for $N=10$
  are practically invisible, making the transfer effectively
  aperiodic. \label{figure7}}
\end{figure}

The beat frequency $\omega_R$ so computed has to be treated with a
grain of salt, however. The reason is that the dynamics in the
radiationless $Q$-space does not preserve probability flux, since the
latter decays into the $P$-space when occupying the modes of outgoing
photons. This is reflected in the presence of damped exponentials in
formula~(\ref{rabi4}). Depending on the number of atoms involved and
the radius of the circle, this damping may be so strong, compared to
the amplitude of the beat oscillations, that the dynamical behaviour
effectively becomes {\it aperiodic}, i.e., exhibits no discernable
oscillations. An example is given by the $N=10$ plot in
Figure~\ref{figure7}.
%----------------------------------------------------------------------
%----------------------------------------------------------------------

\subsection{Analogy between $p=0$ states and hydrogen-like $s$ states}

It is interesting to note that for the $p\neq 0$ states, the central
atom is unoccupied, irrespective of the radius of the circle, or the
number of atoms in the configuration. This means that the central atom
takes part in the dynamics only in a $p=0$ state. This is strongly
reminiscent of the behaviour of scalar single-particle wavefunctions
in a Coulomb potential, such as a spinless electron in a hydrogen
atom: In this case, the electronic wavefunction vanishes at the origin
of the coordinate system, i.e., at the center-of-symmetry of the
potential, for all states with orbital angular momentum quantum number
$l$ greater than zero. On the other hand, in the case of our planar
atomic system, the circular configurations also have a
center-of-symmetry, namely the center of the circle. We can interpret
the ``Frenkel excitons'' $\lst \C{C}_p \Ket / \left\| \C{C}_p
\right\|$ as states of a {\it single quasi-particle}, which is
distributed over the set of discrete locations $\v{R}_z, \v{R}_A$
corresponding to the sites of the atoms. Then the amplitudes $\Bra
A,0| \C{C}_p \Ket= c^p_A$ play a role analogous to a spatial
wavefunction $\Bra \v{x} |\psi\Ket = \psi(\v{x})$; and just as the
hydrogen-like wavefunctions vanish at the origin for angular momentum
quantum numbers $l \neq 0$ \cite{BransdenJoachainBuch2Ed}, so vanish
our quasi-particle wavefunctions at the central atom for all quantum
numbers other than $p=0$. In both cases, the associated wave functions
are {\it isotropic}: The $s$-states transform under the identity
representation ($l=0$) of $SO(3)$ in the case of hydrogen-like
systems, and states $\lst \C{C}_{0(\pm)} \Ket$ under the identity
representation $(p=0)$ of $\ZZ_N$ in the case of our circular
configurations. This means that the quantum number $p$ is analogous to
the angular momentum quantum number $l$ in the central-potential
problem; with hindsight, this may not surprising, since both quantum
numbers $p$ and $l$ are indices which label the unitary irreducible
representations of rotational symmetry groups $\ZZ_N$ and $SO(3)$.

\section{Photon trapping in maximally subradiant states}
\label{PhotonTrapping}

\begin{figure}[b]
\begin{minipage}[b]{0.4\textwidth}
   \centering \includegraphics[width=1.\textwidth]{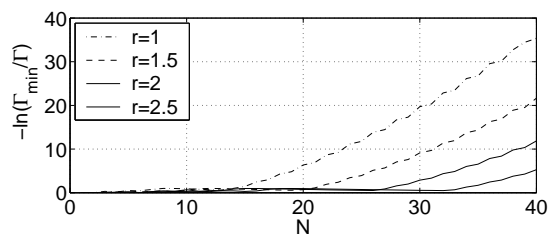}
\end{minipage}
\caption{The negative logarithm of the minimal relative decay rate,
being proportional to the order of magnitude of suppression of
spontaneous decay, as a function of the number of atoms $N$, for fixed
radii as given in the legend. We see that for each radius there exists
a critical number $\hat{N}$ of atoms in the configuration beyond which
the order of magnitude by which spontaneous emission is suppressed is
roughly proportional to $N$. In the Figure, these critical values are
roughly at $\hat{N}=14, 20, 26, 33$ atoms. Beyond these values, the
next-neighbour distance, which is roughly equal to $2\pi r/N$, becomes
smaller than the critical distance $\lambda_{eg}/2$.
\label{figure8}}
\end{figure}

In this section we are interested in the photon-trapping capability of
maximally subradiant states, for large numbers of atoms in the
circle. To this end we choose a fixed radius, increase the number of
atoms in the configuration gradually, and, for each number $N$,
compute the decay rate $\Gamma_{\min}$ of the maximally subradiant
state for the given pair $(r,N)$, in configuration (b) only. We then
expect a more or less monotonic decrease of $\Gamma_{\text{min}}$ as
$N$ increases. But what precisely is the law governing this decrease?
A numerical investigation gives the following result: In
Fig.~\ref{figure8} we plot the negative logarithm
$-\ln(\Gamma_{\text{min}} / \Gamma)$ of the minimal relative decay
rate at the radii $r=1, 1.5, 2, 2.5~\lambda$ for increasing numbers of
atoms. We see that from a certain number $N =\hat{N}$ onwards, which
depends on the radius, $-\ln(\Gamma_{\text{min}} / \Gamma)$ increases
approximately linearly with $N$; in the figure, we have roughly
$\hat{N}=14$ for $r=\lambda$, $\hat{N}=20$ for $r=1.5~\lambda$,
$\hat{N}=26$ for $r=2~\lambda$, $\hat{N}=33$ for $r=2.5~\lambda$. We
also see that the slope is a function of the radius $r$.

We note that the four curves in Fig.~\ref{figure8} imply the existence
of a common {\it critical interatomic distance}: For large $N$, the
next-neighbour distance $R_{nn}$ between two atoms on the perimeter of
the circle is roughly equal to $2\pi r /N$; if we compute this
distance for the pairs of values $(r,\hat{N})$ as found above we
obtain $R_{nn} = 0.45, 0.47, 0.48, 0.48~\lambda$, respectively. We see
that a critical next-neighbour distance of $R_c\approx 0.5~\lambda$
presents itself: If, for fixed radius $r$, the number of atoms in the
configuration is increased, the next-neighbour distance $R_{nn}$
decreases; as soon as $R_{nn} =R_c$ is reached, the order of magnitude
by which spontaneous emission from the maximally subradiant state is
suppressed becomes approximately proportional to $N$.
%----------------------------------------------------------------------
%----------------------------------------------------------------------
Based upon this reasoning we see that, for a given radius $r$, the
critical number $\hat{N}$ of atoms in configuration~(b) is given by
$\hat{N} = \frac{4\pi r}{\lambda}$. Then,  we infer
from Fig.~\ref{figure8} that, approximately,
\begin{equation}
 \Gamma_{\text{min}} \simeq \Gamma \cdot e^{-s(r) (N- \hat{N})} \quad,
 \quad \text{for $N>\hat{N}$} \quad,
\label{ExpLaw1}
\end{equation}
where $s(r)$ determines the slopes of the curves in Fig.~\ref{figure8};
this function decreases monotonically with $r$. We must have the limit
$s(r) \xrightarrow{r\rightarrow\infty} 0$, because for large radii all
correlations between atoms must cease to exist, and hence
$\Gamma_{\text{min}} \rightarrow \Gamma$ in this limit. Let $\tau
\equiv 1/\Gamma$ denote the lifetime of the excited level in a single
two-level atom; then formula (\ref{ExpLaw1}) tells us that the
lifetime $\tau_{\text{min}}$ of a maximally subradiant state
pertaining to $\Gamma_{\text{min}}$ increases exponentially with the
number of atoms,
\begin{equation}
\label{ExpLaw2}
 \tau_{\min} = \tau\cdot e^{s(r) (N- \hat{N})} \quad, \quad \text{for
 $N>\hat{N}$} \quad.
\end{equation}
Theoretically, at least, we can therefore keep a photon trapped in a
circular configuration for arbitrarily large lifetimes, just by
increasing the number of atoms in the circle, hence decreasing the
next-neighbour distance. In practice, of course, a lower limit would
be attained when the wavefunctions of adjacent atoms start to
overlap. Furthermore, the impact of the environment on the coherence
of the maximally subradiant state must be taken into account:

\section{Dephasing: the impact of an environment} \label{Dephasing}

The results presented so far have been idealized in that it was
assumed that the system was closed, being protected from random noise
and loss of quantum coherence into the environment and thus evolving
unitarily. In practice, of course, this cannot be the case any
longer. Through interactions and subsequent entanglement with an
environment, the subsystem atoms+radiation will loose the purity of
the initial state, which will turn into a mixture. This loss is
associated with the intrusion of random elements into the evolution of
the subsystem, of which two generic types may be conceptualized:

Firstly, the sites $\v{R}_A, \v{R}_z$ of the atomic dipoles will
undergo random fluctuations at finite temperature; these fluctuations
will impart small random phases on the radiative contributions emitted
from the individual atoms in the configuration, so that the radiation
from the sample will no longer be completely coherent. This reduction
in coherence will reduce both the super- and the subradiance of
collective states, since it is the interference based on coherence
which is responsible for these effects. We therefore expect that the
radiative decay rates of the superradiant states become smaller, while
those of the subradiant states will increase. Nevertheless, this
effect may be expected to be small, since the amplitudes of
oscillations of the atomic sites will be on the order of a Bohr
radius, while, in the present scenario, the wavelengths of the
electronic transitions are in the optical regime.

Secondly, quantum coherence of the subsystem atoms+radiation will be
reduced by phonon-induced or collisionally-induced dephasing in the
phases of the quantum states, produced by small random shifts in the
unperturbed ground- and excited-state energies of the single atoms due
to the coupling to the environment. It is possible to take these
random elements into account phenomenologically by averaging over the
acquired random phase shifts; in the limit of infinitely short memory
(Markoff approximation) we then arrive at a dephasing-decay rate which
must be added to the radiative decay rate \cite{MeystreSargent}.

In such a case the evolution of the subsystem atoms+radiation must be
described by an appropriate master equation for the associated reduced
density operator. For example, in a solution, the typical nonradiative
decay channel will be comprised of conversions of the exciton energy
into thermal vibrational energy, in other words, a transition from the
electronic excitation into a large number of phonons. The phase of the
coherent, wavelike motion of the exciton, as exemplified in
Fig.~\ref{figure3}, then decays on account of the interaction with
intra- and intermolecular vibrations. These vibrations, in turn, give
rise to fluctuations of the excitation energies and of the interaction
matrix elements and must be taken into account in the Hamiltonian by
an additional, stochastically time-dependent contribution. Averaging
the equation of motion of the density operator over the fluctuations
results in the stochastic Liouville equation describing the exciton
dynamics
\cite{KuboJMP1963a,HakenStroblCUP1967a,HakenReinekerZPh1972a,HakenStroblZPh1973a,AslangulEAPRB1974a,ReinekerPRB1979a,AslangulEAAdChPh1980a}. The
phonons can also be treated as a quantum-mechanical heat bath
\cite{HakenReinekerCUP1968a,WeidlichHaakeZPhys1965a,WeidlichHaakeZPhys1965b,HaakeSpringerverlag1973a}.

In each case additional, nonradiative decay channels open up, with
associated decay rates. Only when these decay rates are small compared
to the idealized radiative decay rates studied in this work will the
results obtained here continue to make sense. We hope to be able to
come back to this point in the future.

\section{Summary}

Collective excitations together with associated level shifts and decay
rates in a planar circular configuration of identical two-level atoms
with parallel atomic dipoles are examined. The relation between these
states to traditional Frenkel excitons is discussed. The state space
of the atomic system can be decomposed into carrier spaces pertaining
to the various irreducible representations of the symmetry group
$\ZZ_N$ of the system. Accordingly, the channel Hamiltonian on the
radiationless subspace of the system can be diagonalized on each
carrier space separately, making an analytic computation of
eigenvectors and eigenvalues feasable. Each eigenvector can be
uniquely labelled by the index $p$ of the associated
representation. For quantum numbers $p >0$ the circular configuration
is insensitive to the presence or absence of a central atom, so that
the wavefunction of the associated quasi-particle describing the
collective excitation occupies the central atom only in a $p=0$
state. It is explained how this feature is analogous to the behaviour
of hydrogen-like $s$-states in a Coulomb potential. The presence of a
central atom causes level splitting and -crossing of the $p=0$ state,
in which case damped oscillations between two "extreme" $p=0$
configurations occur. For strong damping, the population transfer
between the two extreme configurations is effectively
aperiodic. Finally, the behaviour of the minimal decay rate in
maximally subradiant states for varying numbers of atoms in the
configuration is investigated; a critical number of atoms,
corresponding to a next-neighbour distance of $\lambda_{eg}/2$ on the
circle, exists, beyond which the lifetime of the maximally subradiant
state increases exponentially with the number of atoms in the
configuration.

%%%%%%%%%%%%%%%%%%%%%%%%%%%%%%%%%%%%%%%%%%%%%%%%%%%%%%%%%%%%%%%%%%%%%%5

\section{Appendix}

Here we prove a technical result which is used in the main part of the
paper:

Let $R_{1A}$ denote the distance between atoms $1$ and $A$ (i.e.,
outer atoms only). Let $F$ be any function of this distance, $F=
F(R_{1A})$. Let $p$ be an integer. Then
\begin{equation}
\label{append1}
 \sum_{A=2}^N F(R_{1A})\, \sin\left( \frac{2\pi p}{N} (A-1) \right) =
 0 \quad.
\end{equation}
{\it Proof:}
\nopagebreak

The sum $S$ can be written as
\begin{equation}
\label{append2}
 S = \sum_{A=2}^N F(R_{1, N-A+2})\, \sin \left( \frac{2\pi p}{N}
 (N-A+1) \right) \quad.
\end{equation}
The sines are equal to
\begin{equation}
\label{append3}
 \sin \left( \frac{2\pi p}{N} (N-A+1) \right) = - \sin \left(
 \frac{2\pi p}{N} (A-1) \right) \quad,
\end{equation}
while the distances satisfy the equations
\begin{equation}
\label{append4}
 R_{1, N-A+2} = R_{1 A} \quad.
\end{equation}
If this is inserted into (\ref{append2}) we obtain an expression which
is the negative of (\ref{append1}), and as a consequence, $S$ must be
zero. \hfill $\blacksquare$

%%%%%%%%%%%%%%%%%%%%%%%%%%%%%%%%%%%%%%%%%%%%%%%%%%%%%%%%%%%%%%%%%%%%%%%%%%%%%

%
\acknowledgements{Hanno Hammer acknowledges support from EPSRC
grant~GR/86300/01.

%%%%%%%%%%%%%%%%%%%%%%%%%%%%%%%%%%%%%%%%%%%%%%%%%%%%%%%%%%%%%%%%%%%%%%%%%%%%%

%%%%%%%%%%%%%%%%%%%%%%%%%%%%%%%%%%%%%%%%%%%%%%%%%%%%%%%%%%%%%%%%%%%%%%%%%%%%%%5

\end{document}